\RequirePackage{amsmath}
\documentclass[12pt]{iopart}
\pdfoutput=1 
\usepackage{graphicx,cite}
\usepackage{amsmath,amsfonts,amssymb}
\usepackage[urlcolor=blue]{hyperref}
\hypersetup{colorlinks=true,allcolors=blue}
\usepackage{graphicx}
\usepackage[T1]{fontenc}

\newcommand{\E}[1]{\left\langle #1\right\rangle}
\newcommand{\Es}[1]{\left\langle #1\right\rangle_{\rm s}}
\newcommand{\f}[1]{\mathbf{#1}}
\newcommand{\x}{\f x}
\newcommand{\y}{\f y}

\newcommand{\om}{\omega}

\newcommand{\norm}[1]{\left\lvert\left\lvert #1 \right\rvert\right\rvert}

\newcommand{\bsig}{\boldsymbol{\sigma}}
\newcommand{\bom}{{\boldsymbol{\omega}}}

\newcommand{\rj}{\hat{\mathrm j}}
\newcommand{\bj}{\hat{\f j}}
\newcommand{\ps}{p_{\rm s}}
\newcommand{\js}{{\f j}_{\rm s}}

\newcommand{\revj}{{\hat{\f j}}^\ddag}
\newcommand{\comprevj}{{\hat{\mathrm j}}^\ddag}
\newcommand{\integrals}{\hat{\mathcal{I}}}

\renewcommand{\P}[1]{\mathbb{P}\left[#1\right]}
\newcommand{\R}{\mathbb{R}}

\newcommand{\Df}{\mathcal{D}\f f}
\newcommand{\Pf}{P[\f f]}
\newcommand{\idp}{\int\Df\Pf}

\usepackage{color}\usepackage{xcolor}
\colorlet{mylinkcolor}{blue!66!black!80}
\definecolor{grey}{rgb}{0.6,0.6,.6}
\definecolor{darkgrey}{rgb}{0.4,0.4,.4}
\definecolor{darkgreen}{rgb}{0,0.4,0}
\definecolor{lightgreen}{rgb}{0,0.7,0}
\definecolor{darkred}{rgb}{0.5,0,0}

\newcommand{\blue}[1]{{\color{black}#1}}

\begin{document}
\title[Feynman-Kac theory via path integrals and It\^o calculus]{Feynman-Kac theory of 
time-integrated functionals: It\^o versus functional calculus}
\author{Cai Dieball and Alja\v{z} Godec}
\address{Mathematical bioPhysics Group, Max Planck Institute for Multidisciplinary Sciences, 37077 G\"ottingen, Germany}
\ead{agodec@mpinat.mpg.de}

\begin{abstract}
The fluctuations of dynamical functionals such as the empirical density and
current as well as heat, work and generalized currents in stochastic
thermodynamics are usually studied within the Feynman-Kac tilting
formalism, which in the Physics literature is typically derived by
some form of Kramers-Moyal expansion, \blue{or in the Mathematical literature via the Cameron-Martin-Girsanov approach}. Here we derive the Feynman-Kac
theory for general additive dynamical functionals directly via It\^o
calculus and via functional calculus, where the latter results in fact appears
to be new. Using Dyson series we then independently recapitulate recent results
on steady-state (co)variances of general additive dynamical
functionals derived recently in \blue{Dieball and Godec ({2022
    \textit{Phys. Rev. Lett.}~\textbf{129} 140601}) and Dieball and
  Godec ({2022 \textit{Phys. Rev. Res.}~\textbf{4} 033243})}. We hope for
our work to put the different approaches to the statistics of
dynamical functionals employed in the field on a common footing,
\blue{and to illustrate more easily accessible ways to the tilting formalism.}
\end{abstract}

\section{Introduction}
Dynamical functionals \blue{and diverse path-based 
  observables \cite{Derrida1998PRL,Bodineau2004PRL,Lebowitz1999JSP,Garrahan2009JPA,Bodineau2012CMP,Maes2020PR,Jack2015PRL,Vanicat2021},}
such as local and occupation times (also known
as the ``empirical density'')
\cite{Kac1949TAMS,Darling1957TAMS,Aghion2019PRL,Carmi2011PRE,Majumdar2002PRL,Majumdar2002PREa,Majumdar2005CS,Bray2013AP,Bel2005PRL}
as well as diverse time-integrated and time-averaged currents
\cite{Maes2008PA,Touchette2009PR,Kusuoka2009PTRF,Chetrite2013PRL,Chetrite2014AHP,Barato2015JSP,Hoppenau2016NJP,Touchette2018PA,Mallmin2021JPAMT,Monthus2021JSMTE,Dechant2021PRX,Dechant2021PRR,Dieball2022PRL,Dieball2022PRR}
are central to ``time-average statistical mechanics''
\cite{Eli_TA,Burov2011PCCP,Lapolla2020PRR}, large deviation theory (see e.g. \cite{Touchette2009PR,Chetrite2014AHP,Touchette2018PA,Mallmin2021JPAMT,Coghi2021PRE}),
\blue{macroscopic fluctuation theory \cite{Bertini2005PRL,Bertini2015RMP,Derrida2007JSMTE}}, 
and path-wise, stochastic
thermodynamics
\cite{Seifert2018PA,Koyuk2020PRL,Pietzonka2016PRE,Seifert2005PRL,Pigolotti2017PRL,Seifert2012RPP,Dechant2021PRX,Dechant2021PRR}.

Several techniques are available for the study of dynamical
functionals, presumably best known is the
Lie-Trotter-Kato formalism \cite{Darling1957TAMS,Trotter} that was employed by Kac in his seminal
work \cite{Kac1949TAMS}. The techniques typically employed in physics rely on
an analogy to quantum mechanical problems (see e.g. \cite{Majumdar2005CS})
or assume some form of the Kramers-Moyal expansion
\cite{Majumdar2002PRL,Ehrhardt2004PRE,Sabhapandit2006PRE,Bray2013AP} (see also
interesting generalizations to anomalous dynamics
\cite{Carmi2011PRE,Bel2005PRL}).

\blue{
Deriving Feynman-Kac theory  \cite{Kac1949TAMS} of such additive
functionals amounts to obtaining a ``tilted'' generator which
generates the time-evolution of the observables under
consideration. The tilted evolution operator can be obtained using the
Cameron-Martin-Girsanov theorem \cite{Cameron_Martin,Girsanov} ---
a well-known technical theorem often employed in the Mathematical
Physics literature
\cite{Chetrite2013PRL,Chetrite2014AHP,Barato2015JSP}.
}

In this paper we develop the It\^o \cite{Ikeda1981,Lapolla2020PRR} and functional 
calculus \cite{Fox1986PRA,Fox1987JSP} approaches to Feynman-Kac
theory, 
\blue{whereby we a focus on the methodology and accessibility for readers
  that are unfamiliar with the Cameron-Martin-Girsanov approach to ``tilting''}.
\blue{We thereby hope to provide two accessible alternative (but
  equivalent) ways to obtaining the tilted generator. While the It\^o
  approach already exists (see e.g.\ \cite{Lapolla2020PRR} for the
  empirical density), our functional calculus approach is a
  generalization of the pedagogical work of Fox
  \cite{Fox1986PRA,Fox1987JSP} and is aimed towards readers who prefer to
  avoid It\^o calculus.  Since both methods are equivalent they yield
  the same tilted generator. This generator is subsequently used to
  re-derive recent results on the statistics of time-integrated densities
  and currents obtained in
  References~\cite{Dieball2022PRL,Dieball2022PRR} using a different,
  more direct, stochastic calculus approach that avoids tilting. In
  particular, these results illustrate the use of the tilted generator
  to derive the statistics of time-integrated observables for finite times, i.e.\ extending beyond large deviation theory.}

The outline of the paper is as follows. In Sec.~\ref{setup}
we provide the mathematical setup of the problem. In Sec.~\ref{Ito} we
derive the Feynman-Kac equation for a general dynamical functional of
diffusion processes
using It\^o calculus. By generalizing the approach by Fox
\cite{Fox1986PRA,Fox1987JSP} we derive in Sec.~\ref{Fox} the
Feynman-Kac equation using functional calculus. In Sec.~\ref{Dyson} we
apply the formalism to compute steady-state (co)variances of general
dynamical functionals using a Dyson-series approach. We conclude with
a brief perspective.

\section{Tilted Generator}
\blue{In this section, we first introduce the considered stochastic
  dynamics and define what we call ``dynamical functionals''. Subsequently we derive the tilted generator (i.e.\ the
  operator generating the time-evolution of time-integrated
  functionals) based on It\^o calculus, and finally equivalently also via functional calculus.}

\subsection{Set-Up}\label{setup}
We consider overdamped stochastic motion in $d$-dimensional space described by the stochastic differential equation 
\begin{align}
\rmd \x_t=\f F(\x_t)\rmd t+\bsig\rmd \f W_t,
\label{SDE}
\end{align}
where $\rmd \f W_t$ is denotes increment of the Wiener process
\cite{Ikeda1981}. The corresponding diffusion constant is $\f
D=\bsig\bsig^T/2$. For simplicity we stick to additive noise whereas
all present results generalize to multiplicative noise $\f D(\x)$ as
described in \cite{Dieball2022PRR}. In the physics literature
Eq.~\eqref{SDE} is typically written in the form of a Langevin equation
\begin{align}
\dot\x_t&=\f F(\x_t)+\f f(t),\label{LE}
\end{align}
with white noise amplitude $\E{\f f(t)\f f(t')^T}=2\f
D\delta(t-t')$. Comparing the two equations, $\f f(t)$
corresponds to the derivative of $\f W_t$, which however (with
probability one) is not differentiable; more precisely, upon taking $\rmd t\to 0$ one has $\norm{\rmd\f W_t/\rmd t}=\infty$ with probability one, which is why the mathematics literature prefers Eq.~\eqref{SDE}. 

If one describes the system on the level of probability densities
instead of trajectories, the above equations translate to the
Fokker-Planck equation $\partial_tG(\x,t|\x_0)=\hat L(\x)G(\x,t|\x_0)$
with conditional density $G(\x,t|\x_0)$ to be at $\x$ at time $t$
after starting in $\x_0$ and the Fokker-Planck operator \cite{Risken1989,Gardiner1985}
\begin{align}
\hat L(\x)=-\nabla_{\x}\cdot\f F(\x)+\nabla_{\x}\cdot\f D\nabla_{\x}=-\nabla_{\x}\cdot\bj_\x,
\label{Fokker-Planck and current operator} 
\end{align}
where we have defined the current operator $\bj_\x\equiv\f F(\x)-\f
D\nabla_{\x}$. \blue{Note that all differential operators act on all functions to the right, e.g.\ $\nabla_{\x}\cdot\f F(\x)g(\x)=g(\x)\nabla_{\x}\cdot\f F(\x)+\f F(\x)\cdot\nabla_{\x}g(\x)$.} Although the approach presented here is more general,
we restrict our attention to (possibly non-equilibrium) steady states
where the drift $\f F(\x)$ is sufficiently smooth and confining to
assure the existence of a steady-state (invariant) density
$\ps(\x)=\lim_{t\to\infty}G(\x,t|\x_0)$ and steady-state current
$\js(\x)=\bj_\x\ps(\x)$. The special case $\js(\x)=\f 0$ corresponds
to equilibrium steady states. For systems that eventually evolve into a  steady state we can rewrite the current operator as \cite{Dieball2022PRR} \blue{(again the differential operator in $\nabla_\x\ps^{-1}(\x)$ also acts on functions to the right if $\bj_\x$ is applied to a function)}
\begin{align}
\bj_\x=\js(\x)\ps^{-1}(\x)-\f D\ps(\x)\nabla_\x\ps^{-1}(\x).\label{current operator decomposed} 
\end{align}
We will later also restrict the treatment to systems evolving from steady-state initial conditions, i.e.\ the initial condition $\x_{t=0}$ is drawn according to the density $\ps$.

We define the two fundamental additive dynamical functionals---
time-integrated current and density--- as
\begin{align}
{\f J}_t &=\int_{\tau=0}^{\tau=t}U(\x_\tau)\circ\rmd \x_\tau\nonumber\\
\rho_t &=\int_0^t V(\x_\tau)\rmd\tau,\label{definitions} 
\end{align}
with
\blue{differentiable and square-integrable (real-valued)
functions $U,V\colon\R^d\to\R$ and $\circ$ denoting the Stratonovich
integral \cite{Stratonovich1966,Ikeda1981,Kampen1992}}. These objects depend on the whole trajectory
$[\x_\tau]_{0\le\tau\le t}$ and are thus random functionals with
non-trivial statistics. In the following we will derive an equation
for the characteristic function of the joint distribution of
$\x_t,\rho_t,\f J_t$ via a Feynman-Kac approach which will then yield
the moments (including variances and correlations) via a Dyson series.
The formalism was already applied to the time-averaged density
$\rho_t/t$ (under the term of local/occupation time fraction)
\cite{Lapolla2020PRR,Lapolla2018NJP,Kac1949TAMS}.  To do so, we need
to derive a tilted Fokker-Planck equation, which we first do via It\^o
calculus and then, equivalently, via a functional calculus. \blue{Note that
the tilted generator can also be found in the literature on large
deviation theory \cite{Chetrite2014AHP,Barato2015JSP} (in this case obtained via the Feynman-Kac-Girsanov approach)}. 

\subsection{Tilting via It\^o's Lemma}\label{Ito}
We first derive a tilted the Fokker-Planck equation using It\^o calculus. From the It\^o-Stratonovich correction term $\rmd U(\x_\tau)\rmd\x_\tau/2$ and $\rmd\x_\tau\rmd\x_\tau^T=2\f D\rmd\tau$ (where $\f D=\bsig\bsig^T/2$) we obtain from Eqs.~\eqref{SDE} and \eqref{definitions} the increments (curly brackets $\{\nabla\dots\}$ throughout denote that derivatives only act inside brackets)
\begin{align}
\rmd{\f J}_\tau &=U(\x_\tau)\circ\rmd \x_\tau=U(\x_\tau)\rmd \x_\tau+\f D\left\{\nabla_\x U\right\}\!(\x_\tau)\rmd\tau\nonumber\\
\rmd \rho_\tau&=V(\x_\tau)\rmd \tau.\label{differentials} 
\end{align}
We use It\^o's Lemma \cite{Ikeda1981} in $d$ dimensions for a
twice differentiable test function
$f=f(\x_t,\rho_t,\f J_t)$ and Eqs.~\eqref{SDE} and \eqref{differentials}, to obtain
\begin{align}
\rmd f
=&\sum_{i=1}^d\frac{\partial f}{\partial x_i}\rmd x_t^i+\frac{\partial f}{\partial \rho}\rmd \rho_t+\sum_{i=1}^d\frac{\partial f}{\partial J_i}\rmd J_t^i\nonumber\\
&+\frac{1}{2}\sum_{i,j=1}^d\left (\frac{\partial^2 f}{\partial x_i\partial x_j}\rmd x_t^i\rmd x_t^j+\frac{\partial^2 f}{\partial J_i\partial J_j}\rmd J_t^i\rmd J_t^j+2\frac{\partial^2 f}{\partial x_i\partial J_j}\rmd x_t^i\rmd J_t^j\right )\nonumber\nonumber\\
=&[(\nabla_{\x} f)+(\nabla_{\f J} f) U(\x_t)][\f F(\x_t)\rmd t+\bsig\rmd\f W_t]+(\nabla_{\f J} f)\f D\{\nabla_{\x} U\}(\x_t)\rmd t+V(\x_t)\partial_\rho f\rmd t\nonumber\\
&+\left(\nabla_{\x}^T\f D\nabla_{\x}+U(\x_t)^2\nabla_{\f J}^T\f D\nabla_{\f J}+2U(\x_t)\nabla_{\x}^T\f D\nabla_{\f J}\right )f\rmd t.
\end{align}
For the time derivative of $f$ this gives
\begin{align}
\frac{\rmd}{\rmd t}f(\x_t,\rho_t,\f J_t)=&\bigg [\left(\f F+\bsig\frac{\rmd\f W_t}{\rmd t}\right )(\nabla_{\x}+U\nabla_{\f J})+\{\nabla_{\x} U\}\f D\nabla_{\f J}\nonumber\\
&+V\partial_\rho+\nabla_{\x}^T\f D\nabla_{\x}+U^2\nabla_{\f J}^T\f D\nabla_{\f J}+2U\nabla_{\x}^T\f D\nabla_{\f J}\bigg ]f(\x_t,\rho_t,\f J_t).\label{Eq8} 
\end{align}
Following this formalism, we move towards a tilted Fokker-Planck
equation \cite{Kac1949TAMS,Lapolla2020PRR}. Using the conditional
probability density $Q_t(\x,\rho,\f J|\x_0)$ we may write (omitting
$\x$ dependence in $\f F,U,V$ for brevity) the evolution equation for $\langle f(\x_t,\rho_t,\f J_t)\rangle_{\x_0}$, i.e.\ the expected value of $ f(\x_t,\rho_t,\f J_t)$ over the ensemble of paths propagating between $\x_0$ and $\x$ in time $t$. Using Eq.~\eqref{Eq8} and integration by parts, we obtain \blue{(note that non-negative functions $V\ge0$ imply $\rho\ge0$, such that one would restrict the $\rho$-integration to $\int_0^\infty\rmd \rho$ as in Ref.~\cite{Lapolla2020PRR})}
\begin{align}
&\frac{\rmd}{\rmd t}\langle f(\x_t,\rho_t,\f J_t)\rangle_{\x_0}=\int\rmd^d x\int_{\blue{-\infty}}^\infty\rmd \rho\int\rmd^dJ f(\x,\rho,\f J)\partial_t Q_t(\x,\rho,\f J|\x_0)\nonumber\\
&=\int\rmd^d x\int_{\blue{-\infty}}^\infty\rmd \rho\int\rmd^dJ\,Q_t(\x,\rho,\f J|\x_0)\nonumber\\
&\left [\f F(\nabla_{\x}+U\nabla_{\f J})+\{\nabla_{\x} U\}\f D\nabla_{\f J}+V\partial_\rho+\nabla_{\x}^T\f D\nabla_{\x}+U^2\nabla_{\f J}^T\f D\nabla_{\f J}+2U\nabla_{\x}^T\f D\nabla_{\f J}\right ]f(\x,\rho,\f J)\nonumber\\
&=\int\rmd^d x\int_{\blue{-\infty}}^\infty\rmd \rho\int\rmd^dJ\,f(\x,\rho,\f J)\bigg[-\nabla_{\x}\f F-U\f F\nabla_{\f J}-\{\nabla_{\x} U\}\f D\nabla_{\f J}-V\partial_\rho\nonumber\\
&\qquad +\nabla_{\x}^T\f D\nabla_{\x}+U^2\nabla_{\f J}^T\f D\nabla_{\f
      J}+2U\nabla_{\x}^T\f D\nabla_{\f J}\bigg]Q_t(\x,\rho,\f J|\x_0).
  \label{partialI}
\end{align}
Since the test function $f$ is an arbitrary twice differentiable function, the resulting tilted Fokker-Planck equation reads
\begin{align}
\partial_t Q_t(\x,\rho,\f J|\x_0)=\hat{\mathcal{L}}_{\x,\rho,\f
  J}Q_t(\x,\rho,\f J|\x_0),
\label{Titletd_FPE}
\end{align}
with the tilted Fokker-Planck operator \footnote{\blue{For non-negative
functions $V\ge0$  an additional boundary term appears at $\rho=0$ upon partial
integration 
in Eq.~\eqref{partialI}, leading to an extra term 
$-V(\x)\delta(\rho)$ in Eq.~\eqref{Titletd_FPE} that ensures conservation of probability
(see [34]).}}
\begin{align}
\hat{\mathcal{L}}_{\x,\rho,\f J}=&-\nabla_{\x} \cdot \f F(\x)+\nabla_{\x}^T\f D\nabla_{\x}-V(\x)\partial_\rho-U(\x)\f F(\x)\cdot \nabla_{\f J}
\nonumber\\&-\{\nabla_{\x} U(\x)\}^T\f D\nabla_{\f J}+U(\x)^2\nabla_{\f J}^T\f D\nabla_{\f J}+2\nabla_{\f J}^T\f D\nabla_{\x} U(\x)\nonumber\\
=&-[\nabla_\x+U(\x)\nabla_{\f J}]\f F(\x)-V(\x)\partial_\rho
+\left [\nabla_\x+U(\x)\nabla_{\f J}\right]^T\f D\left [\nabla_\x+U(\x)\nabla_{\f J}\right].
\label{tilted_FPE_multidim}
\end{align}
We see that the $\rho$ dependence enters in standard Feynman-Kac form
\cite{Kac1949TAMS,Lapolla2020PRR}, whereas the ${\f J}$ dependence
enters less trivially and shifts the gradient operator $\nabla_\x\to \nabla_\x+U(\x)\nabla_{\f J}$.

\subsection{Tilting via functional calculus}\label{Fox}
We now re-derive the tilted Fokker-Planck operator in
Eq.~\eqref{tilted_FPE_multidim} using a functional calculus approach
\cite{Fox1986PRA,Fox1987JSP} instead of the It\^o calculus in the
previous section. This shows that both alternative approaches are equivalent, as expected. 
We closely follow the derivation of the Fokker-Planck equation in reference
\cite{Fox1986PRA} but for $d$-dimensional space and we generalize the
approach to include the functionals defined in
Eq.~\eqref{definitions}.
\blue{The following approach is equivalent to a Stratonovich
  interpretation of stochastic calculus which is manifested in the
  convention $\int_0^t\delta(t')\rmd t'=\int_0^t\delta(t-t')\rmd
  t'=1/2$ \cite{Fox1986PRA}.}
The white noise term $\f f(\tau)$ with $\Es{\f f(\tau)\f
  f(\tau')^T}=2\f D\delta(\tau-\tau')$ in the Langevin equation
\eqref{LE} can be considered to be described by a path-probability
measure \cite{Fox1986PRA} 
\begin{align}
P[\f f]&=N\exp\left [-\frac{1}{2}\int_0^t\f f(\tau)^{T}\f D^{-1}\f f(\tau)\rmd\tau\right ],\label{Pf}
\end{align}
with normalization constant $N$ which may formally be problematic but always cancels out. 

We now derive a tilted Fokker-Planck equation for the joint conditional density $Q$ of $\x_t$ and the functionals $\f J_t,\rho_t$, as defined in Eq.~\eqref{definitions}, given a deterministic initial condition $\x_0$ at time $t=0$,
\begin{align}
Q_t(\x,\rho,\f J|\x_0)\equiv\idp\delta(\x-\x_t)\delta(\rho-\rho_t)\delta(\f J-\f J_t).
\end{align}
Note for the time derivatives that $\dot{\f J}_t=U(\x_t)\dot{\f x}_t$ and
$\dot{\rho}_t=V(\x_t)$ to obtain (as a generalization of
the calculation in reference \cite{Fox1986PRA} to dynamical functionals)
\begin{align}
\partial_t Q&(\x,\rho,\f J,t|\x_0)=\partial_t \idp\delta(\x-\x_t)\delta(\rho-\rho_t)\delta(\f J-\f J_t)\nonumber\\
&=\idp\left[-\nabla_\x\cdot\dot\x_t-\partial_\rho\dot \rho_t-\nabla_{\f J}\cdot\dot{\f J}_t\right ]\delta(\x-\x_t)\delta(\rho-\rho_t)\delta(\f J-\f J_t)\nonumber\\
&=\idp\left[-\nabla_\x\cdot\left [\f F(\x_t)+\f f(t)\right]-V(\x_t)\partial_\rho-U(\x_t)\left [\f F(\x_t)+\f f(t)\right]\nabla_{\f J}\right ]\times\nonumber\\
&\qquad\delta(\x-\x_t)\delta(\rho-\rho_t)\delta(\f J-\f J_t)\nonumber\\
&=\left[-\nabla_\x\f F(\x)-V(\x)\partial_\rho-U(\x)\f F(\x)\nabla_{\f J}\right]Q_t(\x,\rho,\f J|\x_0)\nonumber\\
&\qquad-\left [\nabla_\x+U(\x)\nabla_{\f J}\right]\cdot\idp\f f(t)\delta(\x-\x_t)\delta(\rho-\rho_t)\delta(\f J-\f J_t).\label{calculation} 
\end{align}
The functional derivative of Eq.~\eqref{Pf} reads \cite{Fox1986PRA}
\begin{align}
\frac{\delta\Pf}{\delta \f f(t)}&=-\frac{1}{2}\f D^{-1}\f f(t)\Pf,\label{deltaPff}
\end{align}
which we use to obtain, via an integration by parts in $\delta \f f(t)$,
\begin{align}
&-\idp\f f(t)\delta(\x-\x_t)\delta(\rho-\rho_t)\delta(\f J-\f J_t)
\nonumber\\&=2\f D\int\Df\frac{\delta\Pf}{\delta \f f(t)}\delta(\x-\x_t)\delta(\rho-\rho_t)\delta(\f J-\f J_t)
\nonumber\\&=-2\f D\idp\frac{\delta}{\delta\f f(t)}\delta(\x-\x_t)\delta(\rho-\rho_t)\delta(\f J-\f J_t).\label{calculation2} 
\end{align}
\blue{As before, differentials are understood to act on all functions to the right, i.e.\ $\frac{\delta}{\delta\f f(t)}$ here acts on the full product of delta functions. We obtain}
\begin{align}
&\frac{\delta}{\delta\f f(t)}\delta(\x-\x_t)\delta(\rho-\rho_t)\delta(\f J-\f J_t)\nonumber\\
&=\left[-\nabla_\x\frac{\delta \x_t}{\delta\f f(t)}-\partial_\rho\frac{\delta \rho_t}{\delta\f f(t)}-\nabla_{\f J}\frac{\delta \x_t}{\delta\f f(t)}\right]\delta(\x-\x_t)\delta(\rho-\rho_t)\delta(\f J-\f J_t),
\end{align}
and we use that ${\delta \rho_t}/{\delta\f f(t)}=\f 0$, and ${\delta \x_t}/{\delta\f f(t)}=\f 1/2$ \cite{Fox1986PRA} which implies ${\delta \f J_t}/{\delta\f f(t)}=U(\x_t)\f 1/2$,
to get
\begin{align}
\frac{\delta}{\delta\f f(t)}\delta(\x-\x_t)\delta(\rho-\rho_t)\delta(\f J-\f J_t)
&=\frac{1}{2}\left[-\nabla_\x-U(\x_t)\nabla_{\f J}\right]\delta(\x-\x_t)\delta(\rho-\rho_t)\delta(\f J-\f J_t).\label{Eq19} 
\end{align}
Plugging Eq.~\eqref{Eq19} first into Eq.~\eqref{calculation2} and then into
Eq.~\eqref{calculation} yields the tilted Fokker-Planck equation for the joint conditional density 
\begin{align}
\partial_t Q_t(\x,\rho,\f J|\x_0)=\Big[&-\nabla_\x\f F(\x)-V(\x)\partial_\rho
-U(\x)\f F(\x)\nabla_{\f J}\nonumber\\
&+\left [\nabla_\x+U(\x)\nabla_{\f J}\right]^T\f D\left [\nabla_\x+U(\x)\nabla_{\f J}\right]\Big]Q_t(\x,\rho,\f J|\x_0).\label{tilted_FPE_multidim_from_functional} 
\end{align}
Note that Eq.~\eqref{tilted_FPE_multidim_from_functional} fully agrees
with Eq.~\eqref{tilted_FPE_multidim} derived via It\^o calculus thus
establishing the announced equivalence of the two approaches.

\section{Steady-state covariance via Dyson expansion of the tilted propagator}\label{Dyson}
\blue{In this section we employ the tilted Fokker-Planck equation
  \eqref{tilted_FPE_multidim_from_functional} to derive results for
  the mean value and (co)variances of time-integrated densities and
  currents. These follow as derivatives of the characteristic function
  evaluated at zero, and it thus suffices to treat the tilt as a perturbation
  of the ``bare'' generator (see \cite{Lapolla2020PRR}).  The derivation is based on a Dyson expansion of the
  exponential of a Fourier-transformed tilted generator (i.e.\ tilted
  Fokker-Planck operator).
  Therefore, consider a one-dimensional Fourier variable $\nu$ and a $d$-dimensional
Fourier variable $\bom=(\om_1,\dots,\om_d)$ and define the
Fourier transform of $Q_t(\x,\rho,\f J|\x_0)$ as
\begin{align}
\tilde Q_t(\x,\nu,\bom|\x_0)\equiv\int_{-\infty}^\infty\rmd \rho\int\rmd^d J\,Q_t(\x,\rho,\f J|\x_0)\exp\left(-\rmi\nu\rho-\rmi\bom\cdot{\f J}\right).
\end{align}
  In the case $V\ge 0$ where $\rho\ge0$ one would instead take the Laplace transform in the $\rho$-coordinate, see Ref.~\cite{Lapolla2020PRR}.}
Recall the \blue{(untilted)} Fokker-Planck operator $\hat L(\x)=-\nabla_\x\cdot\bj_\x$ with the current operator $\bj_\x=\f F(\x)-\f D\nabla_\x$ from Eq.~\eqref{Fokker-Planck and current operator}.
The Fourier 
transform of the tilted Fokker-Planck operator in Eqs.~\eqref{tilted_FPE_multidim} and \eqref{tilted_FPE_multidim_from_functional} reads
\begin{align}
\hat{\mathcal{L}}(\x,\blue{\nu},\bom)&=\hat L(\x)-\blue{\rmi\nu} V(\x)-\rmi \bom^T\cdot \hat{\mathbf{L}}^U(\x)-U(\x)^2\bom^T\f D\bom,\nonumber\\
\hat{\mathbf{L}}^U(\x)&\equiv U(\x) \bj_{\x}-\f D\nabla_{\x} U(\x).\label{Fourier_Generator} 
\end{align}
\blue{As always, the differential operators act on all functions to the right unless written inside curly brackets, i.e.\ $\nabla_{\x} U(\x)=\{\nabla_{\x} U(\x)\}+U(\x)\nabla_{\x}$.
Note that whereas we obtained the tilted generator directly and only
subsequently Fourier transformed it, there are also approaches that directly target the Fourier image of the tilted generator (see e.g.\ \cite{TiznEscamilla2019}).}
Compared to the tilt of the density (i.e. the $\nu$-term; see
also \cite{Lapolla2020PRR}), the tilt corresponding to the current
observable ($\bom$-terms) involves  more terms and even a term that is
second order in $\bom$.  The second order term occurs since $(\rmd\f
W_\tau)^2\sim\rmd\tau$ and therefore (in contrast to $\rmd\tau\rmd\f
W_\tau$ and $\rmd\tau^2$) contributes in the tilting of the
generator. 

We now restrict our attention to dynamics starting in the steady state
$\ps$ and denote the average over an ensemble over paths propagating
from the steady state by $\Es{\cdot}$. Extensions of the formalism to any initial distribution are
straightforward and introduce additional transient terms. \blue{For
  the derivation of the moments of $\rho_t$ and $\f J_t$, we introduce
  and expand} the characteristic function (also known as moment-generating function)
\begin{align}
&\tilde{\mathcal{P}}^{\rho\f J}_t(\blue{\nu},\bom|\ps)\equiv \Es{\rme^{-\blue{\rmi\nu} \rho_t-\rmi\bom\cdot\f J_t}}=1-\blue{\rmi\nu}\Es{\rho_t}-\rmi\bom\cdot\Es{{\f J}_t}\blue{-\nu}\bom\cdot\Es{\rho_t{\f J}_t}+O(\bom^2,\blue{\nu^2}).\label{characteristic function order 2}
\end{align}
\blue{This expansion in $\nu,\bom$ will now be compared to the Dyson expansion of the exponential of Eq.~\eqref{Fourier_Generator} which yields expressions for $\Es{\rho_t},\Es{{\f J}_t},\Es{\rho_t{\f J}_t}$ by comparing individual orders.}

The Dyson expansion allows to expand \blue{for small $|\nu|,|\bom|$}
(see also \cite{Lapolla2020PRR})
\begin{align}
\rme^{\hat{\mathcal{L}}(\x_1,v,\bom)t}=&1\blue{-\rmi}\int_0^t\rmd t_1\rme^{\hat L(\x_1)(t-t_1)}\left [\blue{\nu} V(\x_1)+\bom^T\cdot \hat{\mathbf{L}}^U(\x_1)\right ]\rme^{\hat L(\x_1)t_1}\nonumber\\
&\blue{-}\int_0^t\rmd t_2\int_0^{t_2}\rmd t_1\rme^{\hat L(\x_1)(t-t_2)}\left [\blue{\nu} V(\x_1)+\bom^T\cdot \hat{\mathbf{L}}^U(\x_1)\right ]\rme^{\hat L(\x_1)(t_2-t_1)}\nonumber\\&\left[\blue{\nu}V(\x_1)+\bom^T\cdot \hat{\mathbf{L}}^U(\x_1)\right ]\rme^{\hat L(\x_1)t_1}+O(\bom^2,\blue{\nu^2}).
 \end{align}
Using that the first propagation only differs from $1$ by total derivatives (recall $\hat L(\x)=-\nabla_\x\cdot\bj_\x$), and using for the last propagation term $\rme^{\hat L(\x_1)t_1}\ps(\x_1)=\ps(\x_1)$, we obtain 
\begin{align}
\tilde{\mathcal{P}}^{\rho\f J}_t(v,\bom|\ps)=&\int\rmd^dx_1\,\rme^{\hat{\mathcal{L}}(\x_1,v,\bom)t}\ps(\x_1)\nonumber\\
=&1\blue{-\rmi}\int\rmd^dx_1\int_0^t\rmd t_1\left [\blue{\nu}V(\x_1)+ \bom^T\cdot \hat{\mathbf{L}}^U(\x_1)\right ]\ps(\x_1)\nonumber\\
&\blue{-}\sum_{l,m=1}^d\int\rmd^dx_1\int_0^t\rmd t_2\int_0^{t_2}\rmd t_1\left [\blue{\nu}V(\x_1)+\bom^T\cdot \hat{\mathbf{L}}^U(\x_1)\right ]\rme^{\hat L(\x_1)(t_2-t_1)}\nonumber\\&\left [\blue{\nu}V(\x_1)+\bom^T\cdot \hat{\mathbf{L}}^U(\x_1)\right ]\ps(\x_1)+O(\bom^2,v^2).
\end{align}
We \blue{substitute} the one-step propagation by the conditional
density $G(\x_2,t|\x_1)=\rme^{\hat L(\x_1)t}\delta(\x_2-\x_1)$
\cite{Kampen1992,Pavliotis2014TAM},
\begin{align}
\int\rmd^dx_1 f(\x_1)\rme^{\hat L(\x_1)(t_2-t_1)}g(\x_1)=\int\rmd^dx_1\int\rmd^dx_2 f(\x_2)G(\x_2,t_2-t_1|\x_1)g(\x_1),
\end{align}
which yields
\begin{align}
\tilde{\mathcal{P}}^{\rho\f J}_t(v,\bom|\ps)=&1\blue{-\rmi}\int\rmd^dx_1\int_0^t\rmd t_1\left [\blue{\nu}V(\x_1)+\bom^T\cdot \hat{\mathbf{L}}^U(\x_1)\right ]\ps(\x_1)\nonumber\\
&\blue{-}\int\rmd^dx_1\int\rmd^dx_2\int_0^t\rmd t_2\int_0^{t_2}\rmd t_1
\left [\blue{\nu}V(\x_2)+\bom^T\cdot \hat{\mathbf{L}}^U(\x_2)\right ]\nonumber\\
&G(\x_2,t_2-t_1|\x_1)\left [\blue{\nu}V(\x_1)+\bom^T\cdot \hat{\mathbf{L}}^U(\x_1)\right ]\ps(\x_1)+O(\bom^2,\blue{\nu^2}).\label{Dyson expansion} 
\end{align}
\blue{This concludes the expansion of the exponential of the Fourier transformed tilted generator. Now, by} comparing the definition and expansion of the characteristic function Eq.~\eqref{characteristic function order 2} with the result Eq.~\eqref{Dyson expansion} from the Dyson expansion, we obtain the moments and correlations of the functionals ${\f J}_t=\int_{\tau=0}^{\tau=t}U(\x_\tau)\circ\rmd \x_\tau$ and $\rho_t=\int_0^t V(\x_\tau)\rmd\tau$.

Note that the first moments (i.e.\ the mean values for steady-state initial conditions) can also be obtained directly \cite{Dieball2022PRR,Maes2008PA} but we obtain them here by comparing \blue{the terms of order $\nu$ and $\bom$ in} Eqs.~\eqref{characteristic function order 2} and \eqref{Dyson expansion},
\begin{align}
\Es{\rho_t}&=\int_0^t\rmd t_1\int\rmd^dx_1 V(\x_1)\ps(\x_1)=t\int\rmd^dx_1V(\x_1)\ps(\x_1)\nonumber\\
\Es{\f J_t}&=t\int\rmd^dx_1 [U(\x_1) \bj_{\x_1}-\f D\nabla_{\x_1} U(\x_1)]\ps(\x_1)=t\int\rmd^dx_1 U(\x_1)\js(\x_1),
\end{align}
where $\nabla_{\x_1} U(\x_1)\ps(\x_1)$ vanishes after integration by parts and $\js(\x_1)\equiv\bj_{\x_1}\ps(\x_1)$ is the steady-state current.

By comparing \blue{the terms of order $\nu\bom$ in} Eqs.~\eqref{characteristic function order 2} and \eqref{Dyson expansion} we have for the steady-state expectation $\Es{\f J_t\rho_t}$ that
\begin{align}
\Es{\f J_t \rho_t}&=\int_0^t\rmd t_2\int_0^{t_2}\rmd t_1\int\rmd^dx_1\int\rmd^dx_2
\nonumber\\&\quad
\left [\hat{\mathbf{L}}^{U}(\x_2)G(\x_2,t_2-t_1|x_1)V(\x_1)+V(\x_2)G(\x_2,t_2-t_1|x_1)\mathbf{L}_{U}(\x_1)\right ]\ps(\x_1)\nonumber\\
&=\int_0^t\rmd t_2\int_0^{t_2}\rmd t_1\int\rmd^dx_1\int\rmd^dx_2\Big[U(\x_2) \bj_{\x_2}G(\x_2,t_2-t_1|\x_1)V(\x_1)\nonumber\\
&\quad+V(\x_2)G(\x_2,t_2-t_1|\x_1)[U(\x_1) \bj_{\x_1}-\f D\nabla_{\x_1} U(\x_1)]\Big]\ps(\x_1).\label{cor ansatz}
\end{align}
We note that for any function $f$ the following identity holds 
\begin{align}
\int_0^t\rmd t_2\int_0^{t_2}\rmd t_1f(t_2-t_1)=\int_0^t\rmd t'(t-t') f(t'),
\end{align}
and further introduce the shorthand notation
\begin{align}
\integrals^t_{\x\y}[\cdots]=\int_0^t\rmd t'(t-t')\int\rmd^dx_1\int\rmd^dx_2 U(\x_1)V(\x_2)[\cdots].
\end{align}
Moreover, we define \blue{the joint density} $P_{\y}(\x,t)\equiv G(\x,t|\y)\ps(\y)$ and \blue{following Ref.~\cite{Dieball2022PRR}}
introduce the dual-reversed current operator
$\revj_\x\equiv{\js(\x)}/{\ps(\x)}+\f D\ps(\x)\nabla_\x\ps^{-1}(\x)=-\bj_\x(\js\to-\js)$
. With these notations, using integration by
parts, and by relabeling $\x_1\leftrightarrow\x_2$ in one term, we rewrite Eq.~\eqref{cor ansatz} to obtain for the correlation, \blue{reproducing the main result of Refs.~\cite{Dieball2022PRL,Dieball2022PRR},}
\begin{align}
\Es{\f J_t \rho_t}-\E{\f J_t}\Es{\rho_t}=&\integrals^t_{\x\y}\Big[\bj_{\x_1}P_{\x_2}(\x_1,t')+\js(\x_1)\ps^{-1}(\x_1)P_{\x_1}(\x_2,t')\nonumber\\
&+\f D\ps(\x_1)\nabla_{\x_1}\ps(\x_1)^{-1}P_{\x_1}(\x_2,t')\Big]-\E{\f J_t}\Es{\rho_t}\nonumber\\
=&\integrals^t_{\x\y}\left[\bj_{\x_1}P_{\x_2}(\x_1,t')+\revj_{\x_1}P_{\x_1}(\x_2,t')-2\js(\x_1)\ps(\x_2)\right],\label{cor} 
\end{align}
We will discuss this result below, but first derive analogous results for (co)variances of densities and currents\blue{, respectively}.

Instead of obtaining $\Es{\rho_t^2}$ from the $\blue{\nu^2}$ order in
Eq.~\eqref{Dyson expansion} we here consider a generalization to  two
densities, $\rho_t=\int_0^t V(\x_\tau)\rmd\tau$ and $\rho'_t=\int_0^t
U(\x_\tau)\rmd\tau$.  The \blue{Fourier}-transformed tilted generator in
Eq.~\eqref{Fourier_Generator} with \blue{Fourier} variables $\blue{\nu,\nu'}$
corresponding to $\rho_t,\rho'_t$ is obtained equivalently and gives
$\hat{\mathcal{L}}(\x,v,v')=\hat L(\x)-\blue{\rmi\nu}V(\x)-\blue{\rmi\nu'}U(\x)$. The related
term in the Dyson series (by an adaption of Eq.~\eqref{Dyson
  expansion} including $\blue{\nu'}U$) becomes
$[\blue{\nu}V(\x_2)+\blue{\nu'}U(\x_2)]G(\x_2,t_2-t_1|\x_1)[\blue{\nu}V(\x_1)+\blue{\nu'}U(\x_1)]\ps(\x_1)$
(see also \cite{Lapolla2020PRR}). \blue{By comparison with the
  characteristic function in Eq.~\eqref{characteristic function order 2} including $\rho_t'$, one obtains} the known result \cite{Lapolla2020PRR,Kac1949TAMS},
\begin{align}
\Es{\rho_t\rho'_t}-\Es{\rho_t}\Es{\rho'_t}=\integrals^t_{\x\y}[P_{\x_2}(\x_1,t')+P_{\x_1}(\x_2,t')-2\ps(\x_1)\ps(\x_2)].\label{rho covar}
\end{align}
For $U=V$ this becomes the variance of $\rho_t$ \blue{which can also be obtained from the order $\nu^2$ in Eqs.~\eqref{characteristic function order 2} and \eqref{Dyson expansion}}.

To obtain the current covariance, we accordingly require a tilted generator with two Fourier variables $\bom,\bom'$ corresponding to ${\f J}_t=\int_{\tau=0}^{\tau=t}U(\x_\tau)\circ\rmd \x_\tau$ and ${\f J}'_t=\int_{\tau=0}^{\tau=t}V(\x_\tau)\circ\rmd \x_\tau$, which can, by the same formalism, be derived as
\begin{align}
\hat{\mathcal L}(\x,\bom,\bom')=&\hat L(\x)-\rmi \bom^T\cdot \hat{\mathbf{L}}^U(\x)-\rmi {\bom'}^T\cdot \hat{\mathbf{L}}^V(\x)-U(\x)^2\bom^T\f D\bom-V(\x)^2{\bom'}^T\f D\bom'\nonumber\\
&-2U(\x)V(\x)\bom^T\f D\bom'\nonumber\\
\hat{\mathbf{L}}^{V}(\x)\equiv&V(\x) \bj_{\x}-\f D\nabla_{\x} V(\x).\label{Fourier generator two currents} 
\end{align}
The Dyson series (by adapting Eq.~\eqref{Dyson expansion}) based on $\hat{\mathcal L}(\x,\bom,\bom')$ for two currents $\f J,\f J'$ reads
\begin{align}
&\tilde{\mathcal{P}}^{\f J\f J'}_t(\bom,\bom'|\ps)=\nonumber\\
1&\blue{-}\int\rmd^dx_1\int_0^t\rmd t_1\left[\rmi\bom^T\cdot \hat{\mathbf{L}}^U(\x_1)+\rmi\bom'^T\cdot \hat{\mathbf{L}}^V(\x_1)+2U(\x_1)V(\x_1)\bom^T\f D\bom'\right]\ps(\x_1)\nonumber\\
&+\int\rmd^dx_1\int\rmd^dx_2\int_0^t\rmd t_2\int_0^{t_2}\rmd t_1
\left [\rmi\bom^T\cdot \hat{\mathbf{L}}^U(\x_2)+\rmi\bom'^T\cdot \hat{\mathbf{L}}^V(\x_2)\right ]\nonumber\\
&G(\x_2,t_2-t_1|\x_1)\left [\rmi\bom^T\cdot \hat{\mathbf{L}}^U(\x_1)+\rmi\bom'^T\cdot \hat{\mathbf{L}}^V(\x_1)\right ]\ps(\x_1)+O(\om^2,\om'^2).\label{Dyson expansion two currents}
\end{align}
The expectation value of the product of current components $\Es{J_{t,n} J'_{t,m}}$ is given by the terms that are linear in $\om_n\om'_m$, i.e. (recall $D_{nm}=D_{mn}$)
\begin{align}
&\Es{J_{t,n} J'_{t,m}}=2tD_{nm}\int\rmd^dx_1\, U(\x_1)V(\x_1)\ps(\x_1)+\int_0^t\rmd t'(t-t')\int\rmd^dx_1\int\rmd^dx_2\nonumber\\
&\quad\left[\hat{L}^{U}_n(\x_2)G(\x_2,t'|\x_1)\cdot \hat{L}^{V}_m(\x_1)\ps(\x_1)+\hat{L}^{V}_m(\x_2)G(\x_2,t'|\x_1)\cdot \hat{L}^{U}_n(\x_1)\ps(\x_1)\right].
\label{current covar ansatz} 
\end{align}
We denote by $\hat=$ equality up to gradient terms that vanish upon integration to write 
\begin{align}
&\hat{L}^{U}_n(\x_2)G(\x_2,t'|\x_1)\cdot\hat{L}^{V}_m(\x_1)\ps(\x_1)
\hat=U(\x_2)\rj_{\x_2,n}G(\x_2,t'|\x_1)\times\nonumber\\
&\quad\left [V(\x_1)\js(\x_1)\ps^{-1}(\x_1)-\ps(\x_1)\f D\nabla_{\x_1}\ps(\x_1)^{-1}-\f D\nabla_{\x_1}V(\x_1)\right ]_{m}\ps(\x_1)\nonumber\\
&\hat=U(\x_2)V(\x_1)\rj_{\x_2,n}[\js(\x_1)\ps^{-1}(\x_1)+\ps(\x_1)\f D\nabla_{\x_1}\ps^{-1}(\x_1)]_mG(\x_2,t'|\x_1)\ps(\x_1)\nonumber\\
&=U(\x_2)V(\x_1)\bj_{\x_2,n}\revj_{\x_1,m}P_{\x_1}(\x_2,t).
\end{align}
Inserting this into Eq.~\eqref{current covar ansatz}, and relabeling in one term $\x_1\leftrightarrow\x_2$ we obtain for the $nm$-element of the current covariance matrix
\begin{align}
\Es{J_{t,n} J'_{t,m}}-\Es{J_{t,n}}\Es{J'_{t,m}}=&2tD_{nm}\int\rmd^dx_1\, U(\x_1)V(\x_1)\ps(\x_1)\nonumber\\&
+\integrals^t_{\x\y}\left [\rj_{\x_1,m}\comprevj_{\x_2,n}P_{\x_2}(\x_1,t')+\rj_{\x_2,n}\cdot\comprevj_{\x_1,m}P_{\x_1}(\x_2,t')\right ].\label{current covar} 
\end{align} 
\blue{This reproduces and slightly generalizes the main result of Refs.~\cite{Dieball2022PRL,Dieball2022PRR} where the diagonal elements ($m=n$) of the covariance matrix were derived.}
This result for the current covariance matrix and Eq.~\eqref{cor} for
the current-density correlation are the natural generalizations of the
density-density covariance Eq.~\eqref{rho covar}, as described in
detail in references \cite{Dieball2022PRL,Dieball2022PRR}, with
the additional $2tD_{nm}$-term in Eq.~\eqref{current covar} arising
from the $(\rmd\f W_\tau)^2$ contribution in $J_{t,n} J'_{t,m}$ \blue{manifested in the term $-2U(\x)V(\x)\bom^T\f D\bom'$ in the tilted generator in Eq.~\eqref{Fourier generator two currents}}.
While the density-density covariance Eq.~\eqref{rho covar} only
depends on integration over all paths from $\x_1$ to $\x_2$ (and vice
versa) in time $t'$ via $P_{\x_1}(\x_2,t')$, the current-density
correlation Eq.~\eqref{cor} instead involves
$\bj_{\x_1}P_{\x_2}(\x_1,t')$ and $\revj_{\x_1}P_{\x_1}(\x_2,t')$
which describe currents at the final- and initial-points, respectively
\cite{Dieball2022PRR}. This notion is further extended in the
result Eq.~\eqref{current covar} where
$\rj_{\x_2,n}\comprevj_{\x_1,m}P_{\x_1}(\x_2,t')$ corresponds to 
products of components of displacements along individual trajectories from $\x_1$ to
$\x_2$ \cite{Dieball2022PRL}. 

\section{Conclusion}
We employed a Feynman-Kac approach to derive moments and correlations
of dynamical functionals of diffusive paths --- the time-integrated
densities and currents. We presented two different but equivalent
approaches to tilting the generator ---  It\^o and functional
calculus. 
\blue{These two approaches illustrate how one can freely choose
  between It\^o and functional calculus to derive results on dynamical
  functionals. In particular, both approaches are accessible without
  further technical mathematical concepts such as the
  Cameron-Martin-Girsanov theorem that is often used in the study of
  tilted generators. Our methodological advance thus provides a
  flexible repertoire of easily accessible methods that will hopefully prove useful in future studies of related problems. 

The derivation of the moments and correlations based on the tilted generator reproduces results with important implications for stochastic thermodynamics and large deviation theory, in particular for the physical and
mathematical role of coarse graining \cite{Dieball2022PRL,Dieball2022PRR}, and thereby displays how the tilted generator yields results on the statistics of dynamical functionals, even beyond the large deviation limit.}

\section*{Acknowledgments}
Financial support from Studienstiftung des Deutschen Volkes (to
C.\ D.) and the German Research Foundation (DFG) through the Emmy
Noether Program GO 2762/1-2 (to A.\ G.) is gratefully acknowledged.

\section*{Data availability statement}
No new data were created or analysed in this study.

\section*{References}
\bibliography{bib_JPhysA.bib}
\end{document}